\documentclass[aps,nofootinbib,groupedaddress]{revtex4}

\usepackage{url}
\usepackage{graphicx}
\usepackage{amsmath,amssymb,amstext,amssymb,amsfonts,amsthm}
\usepackage{hyperref}
\usepackage{appendix}
\usepackage[margin=1.0in,papersize={8.5in,11in}]{geometry}
\usepackage[utf8]{inputenc}
\usepackage{soul}
\usepackage{color}

\usepackage{tabularx}
\newcolumntype{C}{>{\centering\arraybackslash}X}
\newcolumntype{R}{>{\raggedleft\arraybackslash}X}

\def\eqref#1{(\ref{#1})}

\begin{document}

\title{Towards detection of relativistic effects in galaxy number counts using kSZ Tomography}

\author{Dagoberto Contreras${}^{1,2}$}
\author{Matthew C. Johnson${}^{1,2}$}
\author{James B. Mertens${}^{1,2,3}$}
\email[]{mertens@yorku.ca}

\affiliation{${}^1$Department of Physics and Astronomy, York University, Toronto, Ontario, M3J 1P3, Canada}
\affiliation{${}^2$Perimeter Institute for Theoretical Physics, Waterloo, Ontario N2L 2Y5, Canada}
\affiliation{${}^3$Canadian Institute for Theoretical Astrophysics, University of Toronto, Toronto, ON M5H 3H8 Canada}

\begin{abstract}
High-resolution, low-noise observations of the cosmic microwave background (CMB) planned for the near-future will enable
new cosmological probes based on re-scattered CMB photons -- the secondary CMB. At the same time, enormous
galaxy surveys will map out huge volumes of the observable Universe. Using the technique of kinetic Sunyaev Zel'dovich
(kSZ) tomography these new probes can be combined to reconstruct the remote dipole field, the CMB dipole as
observed from different vantage points in our Universe. The volume accessible to future galaxy surveys is large enough that general
relativistic corrections to the observed distribution of galaxies must be taken into account. These corrections are
interesting probes of gravity in their own right, but can also obscure potential signatures of primordial non-Gaussianity.
In this paper, we demonstrate that correlations between the reconstructed remote dipole field and the observed galaxy
number counts can in principle be used to detect general relativistic corrections. We show that neglecting
general relativistic corrections leads to an $\mathcal{O}(1)$ bias on the inferred amplitude of primordial non-Gaussianity, $f_{\rm NL}$.
In addition, we demonstrate that the reconstructed remote dipole field can provide useful constraining power on various
bias parameters appearing in the galaxy number counts, and can significantly mitigate the effects of alignment bias.
\end{abstract}

\maketitle

\section{Introduction}

Galaxy surveys and cosmic microwave background (CMB) measurements will provide us with exceptionally
accurate and precise measurements of our Universe over the coming decade.
Galaxy surveys such as LSST~\cite{0912.0201} will produce massive redshift catalogues on the volume frontier,
mapping out structures on ultra-large scales. Several exciting opportunities present themselves in
the era of large-volume surveys, including the potential to measure subtle general relativistic effects in
 the observed clustering of galaxies~\cite{Sasaki:1987ad,Pyne:2003bn,Hui:2005nm,Bonvin:2005ps,Barausse:2005nf,Yoo:2008tj,2009PhRvD..80h3514Y,Yoo:2010ni,1105.5280,1105.5292,1107.5427}
and an opportunity to detect primordial non-Gaussianity through its scale-dependent effect
on galaxy bias~\cite{Dalal:2007cu}. CMB experiments such as Simons Observatory~\cite{Ade:2018sbj}
and CMB-S4~\cite{1610.02743} will produce measurements of the CMB on the sensitivity frontier,
mapping out small-scale CMB temperature and polarization anisotropies near the nano-Kelvin level.
At this sensitivity, it will be possible to accurately measure secondary CMB anisotropies such as the kinetic
Sunyaev Zel'dovich (kSZ) effect~\cite{SZ80}, temperature anisotropies induced by the scattering of CMB
photons from free electrons in bulk motion after reionization.

At first sight, these developments might appear only vaguely related. However, because the large scale structure (LSS) is
responsible for the secondary CMB anisotropies, there is a strong correlation between the small-angular scale CMB
and the distribution of structure probed by e.g. galaxy surveys. In the case of the kSZ effect, the statistical anisotropies
in such correlations encode information about the structure of the Universe on the largest scales. This information can
be extracted using a technique known as kSZ tomography~\cite{Ho09,Shao11b,
Zhang11b, Zhang01,Munshi:2015anr,Schaan15,Ferraro:2016ymw,Hill:2016dta,Zhang10d,Zhang:2015uta,Terrana2016,Deutsch:2017ybc,Smith:2018bpn,Munchmeyer:2018eey,Sehgal:2019nmk},
described in more detail below. In this paper, we demonstrate that a comparison of the large-scale distribution
of galaxies to the large-scale inhomogeneities reconstructed using kSZ tomography can yield valuable new information
about general relativistic contributions to the observed distribution of galaxies, with consequences for the
measurement of primordial non-Gaussianity and various astrophysical bias parameters.

Considering the ways in which relativistic effects contribute to observables
dates back to the birth of general relativity. In the present context, such effects arise from a
precise treatment of photon geodesics in an inhomogeneous Universe. On large scales, cosmological
perturbation theory is formulated within general relativity (GR), so there are no additional
``dynamical'' effects to consider. The earliest perturbative calculations that carefully considered
all relativistic contributions to observables for scalar perturbations
are perhaps for fluctuations in the luminosity distance-redshift relation~\cite{Sasaki:1987ad,Pyne:2003bn,Hui:2005nm,Bonvin:2005ps}.
Considering these effects will be important for future supernova surveys to constrain the properties
of dark energy~\cite{Hui:2005nm}. Subsequently, the importance of relativistic effects has also been
considered beyond leading order \cite{Barausse:2005nf,1205.5221,1207.1286,1207.2109,1401.7973,1402.1933,1405.7860,1406.1135,1406.4140,1812.04336}, and in an exact, numerical
setting \cite{Giblin:2016mjp,Giblin:2017ezj}.

The galaxy number density as a function of redshift and angle on the sky--what we actually observe in a galaxy redshift survey--is also
subject to relativistic corrections~\cite{Yoo:2008tj,2009PhRvD..80h3514Y,Yoo:2010ni,1105.5280,1105.5292,1107.5427},
similar in nature to effects commonly considered for other observables such as the CMB.
In addition to the projected galaxy number density, redshift-space distortions (RSD; anisotropies in
the mapping from real-space to redshift-space induced by peculiar velocities), magnification from lensing,
and subdominant ``general relativistic'' (GR) effects also contribute to the
observed number counts at linear order.\footnote{Rigorous treatments of relativistic
effects have been extended to second order in~\cite{Bertacca:2014dra,Bertacca:2014wga}.}
These GR effects include additional Doppler (magnification) terms
and potential (Sachs-Wolfe, integrated Sachs-Wolfe, time delay) terms, analogous to the CMB.
GR effects on number counts become important on scales approaching the cosmological horizon, and
therefore are only accessible to surveys which have very large volume. Using a single tracer, it is
unlikely that GR effects can be detected by any near-term survey~\cite{1505.07596,1710.02477}. However, multi-tracer
techniques can be used to pull them out~\cite{Alonso:2015sfa}. Should these effects be detectable, they could serve
as a probe of modifications to GR on ultra-large scales~\cite{Baker:2015bva,Renk:2016olm,Duniya:2019mpr}.

A prime science target of future galaxy surveys, for example LSST~\cite{0912.0201} and SPHEREX~\cite{Dore:2014cca}, is primordial non-Gaussianity (PNG);
see~\cite{Alvarez:2014vva} for an overview. Local-type PNG induces a scale-dependent galaxy
bias~\cite{Dalal:2007cu}, proportional to the parameter $f_{\rm NL}$, leading to a measurable enhancement ($f_{\rm NL}>0$) or suppression ($f_{\rm NL}<0$)
of galaxy clustering on the largest scales. An important milestone is constraining PNG at the level of
$\sigma(f_{\rm NL}) < 1$. This is because a generic prediction of a large class of multi-field inflationary models
is $f_{\rm NL} \sim \mathcal{O}(1)$, making a constraint at this level a potentially powerful discriminator between
single-field and multi-field models of inflation~\cite{Bartolo:2004if,Alvarez:2014vva}.
Both PNG and GR effects modify the galaxy-galaxy power spectrum on large physical/angular scales,
making it important to incorporate both into forecasts for the constraining power of future experiments. In particular, it is known that neglecting
GR effects can lead to an $\mathcal{O}(1)$ bias on measurements of $f_{\rm NL}$~\cite{1710.02477}. A proper
understanding of GR effects is therefore essential for properly interpreting  future surveys, and the implication of
their results for inflationary cosmology.

Measurements of PNG~\cite{Seljak:2008xr} and GR effects~\cite{Alonso:2015sfa} can benefit greatly from incorporating
cross-correlations of galaxy surveys with other tracers of the underlying dark matter distribution. In this case, a mode-by-mode
comparison between the galaxy number density and dark matter density can be used to measure (scale-dependent) bias or
to identify extra contributions to the observed galaxy number density from GR effects. Because such a comparison depends
only on the properties of our observed realization, there is in principle no sample variance~\cite{Seljak:2008xr}. Taking advantage
this ``sample variance cancellation'', the ability to measure PNG or GR effects is limited only by the fidelity of the
reconstruction of the dark matter density field (which depends on how correlated the tracer is with the distribution of dark matter)
and shot noise on the galaxy survey.

Recently, kSZ tomography was introduced as a new and powerful tool for constraining PNG~\cite{Munchmeyer:2018eey}.
The kSZ effect induces CMB temperature anisotropies due to the scattering of CMB photons from free electrons in the
post-reionization Universe. The kSZ effect comprises the dominant blackbody contribution to the observed CMB temperature
on small angular scales ($\ell \gtrsim 4000$). A number of detections of the
kSZ effect have been made with existing
datasets~\cite{Hand12,Schaan15,PlanckkszI,DeBernardis2016pdv,Soergel2016mce,Sugiyama2017uvr,Li2017uin,Hill2016dta,PlanckkszII},
and future experiments promise to obtain very high significance measurements~\cite{Ade:2018sbj,1610.02743}.
The amplitude of the kSZ temperature anisotropy from locations along our past light cone is proportional to the value of the remote dipole field,
the projected temperature dipole at different locations in the observable Universe.

The three-dimensional remote dipole field can be reconstructed from statistical anisotropies in the cross-correlation
between the CMB temperature on small angular scales and a galaxy survey, a technique called kSZ tomography~\cite{Terrana2016,Deutsch:2017ybc,Smith:2018bpn}.
The dominant contribution to the remote dipole field is the local peculiar velocity, which can be related to the
dark matter density through linear theory. Correlating the reconstructed dipole field with the galaxy survey can
therefore be used to isolate the scale-dependent galaxy bias due to PNG; the high fidelity of the reconstruction
possible with future surveys allows one to take strong advantage of sample variance cancellation. Ref.~\cite{Munchmeyer:2018eey}
forecasted that it will in principle be possible to constrain PNG at the level of $\sigma(f_{\rm NL}) < 1$ with the
next generation of CMB instruments and galaxy surveys using kSZ tomography.

The present paper makes a number of important contributions to this previous analysis. First, we extend the analysis of Ref.~\cite{Munchmeyer:2018eey}, which
utilized a simplified geometry, to the light cone. Our analysis includes all contributions to the remote dipole
field beyond the local peculiar velocity (Doppler, Sachs-Wolfe, and integrated Sachs-Wolfe). We include RSD, lensing, and GR effects
in the galaxy number density. We leave as free parameters in our model the redshift-dependent galaxy bias, evolution bias, magnification bias, and a multiplicative
bias on the reconstructed remote dipole field that describes the optical depth degeneracy (see e.g.~\cite{Hall2014,Battaglia:2016xbi}) in measurements of the kSZ effect
(see Ref.~\cite{Smith:2018bpn} for an argument that this is sufficient). We
also include an additional bias in the galaxy survey
from intrinsic alignments due to large-scale tidal fields~\cite{0903.4929}.
From here on we refer to such a bias as the alignment bias, defined explicitly
in~\cite{0903.4929} and Appendix~\ref{appendix:numbercounts}. Finally, we include information from the primary CMB temperature and polarization in our constraints.

The goals of the present paper are to answer the questions:
\begin{itemize}
\item Is it possible to isolate GR effects in galaxy surveys using kSZ tomography?
\item To what extent does neglecting GR effects bias the measurement of $f_{\rm NL}$ using kSZ tomography?
\item Is sample variance cancellation between the remote dipole field and the galaxy survey useful for measuring various bias parameters?
\item Does incorporating information from the primary CMB temperature and polarization help?
\end{itemize}
In summary, we find that kSZ tomography is a useful tool for isolating GR effects and improving the measurement of
a variety of bias parameters when information from the primary CMB is incorporated. We further demonstrate that neglecting GR effects leads to an $\mathcal{O}(1)$
bias on measurements of $f_{\rm NL}$ from kSZ tomography.

The paper is structured as follows. In Sec.~\ref{sec:forecasting_method}, we outline the parameters of the Fisher forecast. In Sec.~\ref{sec:results},
we discuss the results of our forecast, and in Sec.~\ref{sec:discussion} we conclude. A number of appendices are included to
summarize the observables which go into our forecast.

\section{Forecasting method}\label{sec:forecasting_method}

In order to address the questions posed in the introduction, we forecast the
constraining power of future CMB measurements and galaxy surveys. Performing this
forecast will require two main ingredients: one or more observable quantities,
and any noise associated with measuring these observables.
The observables we consider are angular power spectra,
\begin{equation}
\label{eq:powerspec}
C_{\ell}^{XY}=4\pi\int\frac{dk}{k}\mathcal{P}(k)\Delta_{\ell}^{X}(k)\Delta_{\ell}^{Y*}(k)\,,
\end{equation}
where $X$ and $Y$ are one of: the perturbations in number counts of galaxies,
the primary CMB temperature and polarization perturbations, or the remote
dipole field reconstructed using kSZ tomography.
Here, $\mathcal{P}(k)$ is the dimensionless primordial power spectrum,
\begin{equation}
\mathcal{P}(k)=A_{s}\left(\frac{k}{k_{0}}\right)^{n_{s}-1}\,.
\end{equation}
The noise associated with each of these observables will be shot noise in the
case of galaxy number counts, instrument noise in the case of the primary CMB,
and reconstruction noise in the case of the remote dipole field.
The angular power spectra themselves are computed at linear order for the different
tracers we consider. We explicitly state the form of the contributions to the
transfer functions for number counts and the remote dipole field in
Appendices~\ref{appendix:numbercounts} and \ref{appendix:ksz}. The noise models we
use are partly described below, and further details appear in these appendices,
along with details on various bias functions that the transfer functions depend on.

As we are interested in exploring the importance of relativistic, non-Newtonian
corrections, we parametrize the amplitude of these effects following \cite{1505.07596}.
We describe the amplitude of the relativistic corrections by defining the parameters
$\epsilon^{\rm N}_{\rm GR}$ and $\epsilon^{\rm kSZ}_{\rm GR}$, defined at the level of the transfer functions
as
\begin{align}
  \Delta_{\ell}^{\rm N}(k) = & \Delta_{\ell}^{\rm D}(k) + \Delta_{\ell}^{\rm RSD}(k) + \Delta_{\ell}^{\rm L}(k)
    + \epsilon^{\rm N}_{\rm GR} \Delta_{\ell}^{\rm N, GR}(k) \\
  \Delta_{\ell}^{\rm kSZ}(k) = & \Delta_{\ell}^{\rm LD}(k) + \epsilon^{\rm kSZ}_{\rm GR} \Delta_{\ell}^{\rm kSZ, GR}(k)\,.
\end{align}
Here, the relativistic terms for number counts ($\Delta^{\rm N, GR}_\ell$) include all effects except standard
intrinsic density fluctuations ($\Delta^{\rm D}_\ell$), RSD ($\Delta^{\rm
RSD}_\ell$), and lensing terms ($\Delta^{\rm L}_\ell$), all of which are
defined in Appendix~\ref{appendix:numbercounts}. While lensing itself is a
general relativistic effect that must be taken into account, it has been considered separately in previous
literature, and we follow this convention. The relativistic contributions
to the remote dipole field ($\Delta^{\rm kSZ, GR}_\ell$) we take to include all contributions except the
local (Newtonian) peculiar velocity Doppler term ($\Delta^{\rm LD}_\ell$). These contributions are also
``primordial'' in the sense that they represent the Sachs-Wolfe, Integrated Sachs-Wolfe,
and primordial Doppler components.

We compute the galaxy number counts spectrum and reconstruct the remote dipole field
in a set of redshift bins between $z \in [0.1, 3]$.\footnote{kSZ tomography can in principle be performed at higher redshifts,
however the number of observed galaxies beyond $z \sim 3$ in our fiducial survey is quite small, leading
to a large reconstruction noise on the remote dipole field.}
We divide this range into 30 bins equally spaced in comoving distance,
and integrate the transfer functions over the redshift range within each bin weighted by
tophat window functions (eg. $W_i(z)$ in Eq.~\ref{eq:ngr_windowfn} is a tophat).
The width of these bins roughly corresponds to photometric redshift uncertainties,
with $\sigma_{z} \sim 3 / 30 \sim 0.1$ in our binning scheme, although a more optimistic
forecast could include additional bins as $\sigma_z$ is projected to be of order 0.05 for
an LSST-like survey, and as low as $\sigma_z \sim 0.02$ for a ``red'' galaxy population~\cite{0912.0201}.
However, generally do not find that changing the number of bins strongly affects our
constraints, as we illustrate further below.

\begin{table}[h]
\renewcommand{\arraystretch}{1.75}
\begin{tabularx}{0.99\textwidth}{rCCCCCCCCCCCCCC}
\hline
\hline
Parameter & $\epsilon_{{\rm GR}}^{{\rm kSZ}}$ & $\epsilon_{{\rm GR}}^{{\rm N}}$
& $f_{NL}$ & $10^{10}A_{s}$ & $n_{s}$ & $\Omega_{b}$ & $\Omega_{c}$ & $h$ & $\tau$ & $b_{v}^{z}$ & $b_{\rm G}^{z}$ & $b_{\rm A}^z$ & $f_{\rm evo}^{z}$ & $s^{z}$ \\
\hline
Fiducial value  & 1 & 1 & 0 & 2.2 & 0.96 & 0.0528 & 0.2647 & 0.675 & 0.06 & 1 & $\dagger$ & 0 & $\dagger$ & $\dagger$ \\
\hline
\end{tabularx}
\caption{Cosmological and relativistic parameters, bias functions, and their
fiducial values. The biases $b_v,\, b_G,\, b_A,\, f_{\rm evo},\, s$, that we refer to
throughout are, respectively, the optical depth bias, the galaxy bias, the
alignment bias, the evolution bias, and the magnification bias all defined
explicitly in Appendix~\ref{appendix:numbercounts} and~\ref{appendix:ksz}. The values of bias functions indicated with a $\dagger$ vary
with redshift, the modeling of which is described in Appendix~\ref{appendix:numbercounts}. }
\label{table:params_list}
\end{table}

The final parameters we consider include standard cosmological ones, $f_{NL}$,
bias parameters in each redshift bin, and the lightcone-projection/GR/primordial
correction coefficients. In Table~\ref{table:params_list} we list each of the parameters we constrain,
along with the fiducial values we use in our forecast.

We compute the Fisher matrix,
\begin{equation}
\label{eq:fisherM}
F_{\alpha\beta} = \sum_{\ell=1}^{\ell_{{\rm max}}}\frac{2\ell+1}{2}{\rm Tr}\left[\left(\partial_{\alpha}\mathcal{C}_{\ell}\right)\mathcal{C}_{\ell}^{-1}\left(\partial_{\beta}\mathcal{C}_{\ell}\right)\mathcal{C}_{\ell}^{-1}\right] + F^{\rm CMB,\,high-\ell}_{\alpha\beta}\,,
\end{equation}
for parameters $\alpha$ and $\beta$, and using a covariance matrix that includes
all of the tracers,
\begin{equation}
\label{eq:covM}
\mathcal{C}_{\ell}=\left(\begin{array}{ccc}
C_{\ell}^{{\rm N,N}} & C_{\ell}^{{\rm N,CMB}} & C_{\ell}^{{\rm N,kSZ}} \\
C_{\ell}^{{\rm CMB,N}} & C_{\ell}^{{\rm CMB,CMB}} & C_{\ell}^{{\rm CMB,kSZ}} \\
C_{\ell}^{{\rm kSZ,N}} & C_{\ell}^{{\rm kSZ,CMB}} & C_{\ell}^{{\rm kSZ,kSZ}} \\
\end{array}\right) + N_{\ell}\,.
\end{equation}
The $C_{\ell}^{XY}$ functions are the power spectra as defined in
Eq.~\eqref{eq:powerspec} and $N_{\ell}$ noise sources associated with
these measurements, described in more detail below. These spectra are computed
in each redshift bin for the remote dipole field and number counts; cross-spectra between all bins
are accounted for, although this is not made explicit in Eq.~\eqref{eq:covM}. The CMB
spectra are computed for both temperature and $E$-mode polarization.
Derivatives are computed numerically with respect to the parameters
$\{\Omega_{b},\Omega_{c},h,\tau\}$ using a second-order accurate
upwind finite difference stencil. The remaining derivatives are computed
analytically, explicitly commuting derivatives through any integrals over the
redshift bins.

Due to the contributions from relativistic effects and non-Gaussianities manifesting
on large scales, we consider auto- and cross-spectra of all tracers
included in the forecast at low $\ell < 60$. We also account for the high-$\ell$ CMB
temperature and $E$-mode constraints on cosmological parameters separately,
using lensed CMB power spectra generated using  \textsc{CAMB}. The Fisher matrix from
this is included as the $F^{\rm CMB,\,high-\ell}_{\alpha\beta}$ term in Eq.~\eqref{eq:fisherM}.
In producing the high-$\ell$ CMB constraints, we have assumed a maximum available
$\ell$ of 4000 in both $T$ and $E$. The high-$\ell$ constraint is also only used to
constrain the standard cosmological parameters $\{A_s, n_s,\Omega_{b},\Omega_{c},h,\tau\}$,
and so elements are considered zero when an index corresponds to the remaining parameters
and bias functions.

In order to study constraints from specific tracers or combinations of
tracers, we can selectively exclude tracers from the covariance matrix $C_{\ell}$ by
removing the row and column associated with a particular
tracer, eg. removing the last column and row in order to neglect the contribution from
the remote dipole field. When excluding the CMB, we also exclude the high-$\ell$ constraint (the $F^{\rm CMB,\,high-\ell}_{\alpha\beta}$ term).
In this way, we can examine how
constraints from number counts alone improve as additional information is added from the
primary CMB, and subsequently the remote dipole field.

The noise spectra $N_{\ell}$ are computed consistently for each observable we consider.
For the primary CMB, we assume a CMB instrument noise of $1 \mu$k-arcmin
for both temperature and polarization measurements (although we vary this later),
and a 1 arcminute beam in each. Galaxy shot noise is computed from the luminosity function
and limiting magnitude of the survey, as described in Appendix~\ref{appendix:numbercounts}.
This is equivalent to the model adopted in Ref.~\cite{1505.07596}.
The reconstruction noise for the remote dipole field is then computed following \cite{Deutsch:2017ybc},
using the galaxy number counts and CMB power spectra and noise described previously. We further assume
that the electron distribution follows the dark matter distribution for our fiducial model. The uncertainty in
this assumption is folded into the optical depth bias on the remote dipole, which we marginalize over in our
analysis. For the reconstruction noise we assume a maximum available $\ell$ of $\ell_{\rm max} = 9000$,
which largely saturates the signal-to-noise of the relevant modes at the assumed CMB noise levels.
We also assume that foregrounds and systematics can be mitigated and we do not
consider these here. This may have an effect on the realistically accessible $\ell_{\rm max}$, and
we comment on the implications for our constraints below.

We lastly seek to evaluate the bias in parameters, $\Delta \alpha$, due
to neglecting various terms that arise from relativistic considerations.
Following \cite{1710.02477}, we have
\begin{equation}
\Delta \alpha = \left(F^{-1}\right)^{\alpha\beta}v_{\beta}\,,
\end{equation}
where $F^{-1}$ is the inverse fisher matrix, and
\begin{equation}
v_{\beta}=\sum_{\ell=1}^{\ell_{{\rm max}}}\frac{2\ell+1}{2}{\rm Tr}\left[\left(\partial_{\alpha}\mathcal{C}_{\ell}\right)\mathcal{C}_{\ell}^{-1}\Delta \mathcal{C}_{\ell} \mathcal{C}_{\ell}^{-1}\right]\,,
\end{equation}
where $\Delta \mathcal{C}_{\ell} = \mathcal{C}_{\ell}^{\rm true} - C_{\ell}^{\rm fiducial}$. The quantity
$\Delta \alpha$ then describes the extent to which the measurement of a parameter $\alpha$ is
biased when assuming a fiducial spectrum $C_{\ell}^{{\rm fiducial}}$ instead of using the true spectrum $C_{\ell}^{{\rm true}}$. The fiducial spectrum we use here
is the one where the general relativistic contributions are neglected in both
the number counts and kSZ measurements, ie. the transfer functions are evaluated with $\epsilon_{{\rm GR}}^{{\rm kSZ}} = 0$
and $\epsilon_{{\rm GR}}^{{\rm N}} = 0$.

\section{Results}\label{sec:results}

\begin{figure}[htb]
  \centering
    \includegraphics[width=0.99\textwidth]{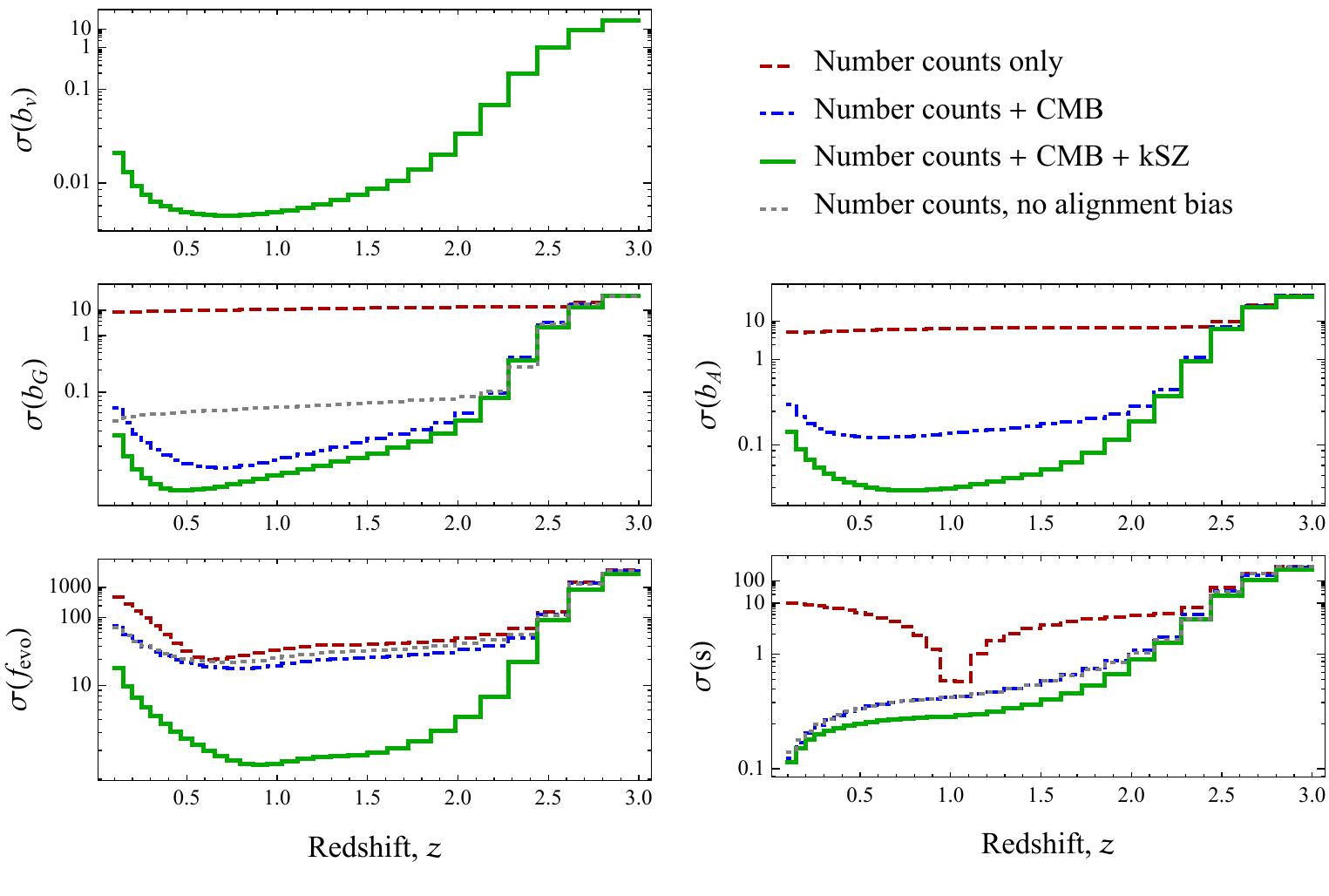}
    \caption{
      Constraints on various biases as a function of redshift bin, including different combinations of tracers as indicated by the legend. Constraints without accounting for alignment bias are also shown. ``Steps'' indicate the width of the redshift bins.
    }
  \label{fig:bias_constraints}
\end{figure}

We now examine the impact of the relativistic effects on future observations of galaxy number counts,
 CMB temperature and polarization, and the reconstruction of the remote dipole field.
We wish to highlight the importance of alignment bias in the context of number
counts alone. We explicitly focus on this as alignment bias is often not taken
into account in forecasts, yet has the potential to interfere with measurements
of the growth function and galaxy bias, thereby degrading constraints on
cosmological parameters. We then incorporate both primary CMB measurements,
and the remote dipole field into the forecast,
and explore how constraints change as each new observable is included. We will
demonstrate that the degradation on parameter constraints incurred by including the
alignment bias is no longer an issue once CMB and kSZ tomography are included.
In particular constraints on $f_{\rm evo}$ are dramatically improved when kSZ
tomography is used (for $z \lesssim 2.5$).

\begin{table}[tb]
\renewcommand{\arraystretch}{1.75}
\begin{tabularx}{0.99\textwidth}{rCCCCC}
\hline
\hline
 & N & N, no $b_A$ & N + CMB & N + CMB + kSZ & { N + CMB + kSZ + $f_{\rm evo}$ prior } \\
\hline
$\sigma(f_{\rm NL})$ & 16 & 14 & 9.0 & 1.3 & 1.0 \\
$b ( f_{\rm NL} )$ & -9.3 & -8.3 & -2.2 & -2.0 & -3.3 \\
$\sigma(\epsilon_{\rm GR}^{\rm N})$ & 5.1 & 4.1 & 3.2 & 0.58 & 0.09 \\
\hspace{1.5em}$\sigma(\epsilon_{\rm GR}^{\rm kSZ})$ & -- & -- & -- & 0.61 &
0.60 \\
\hline
\end{tabularx}
\caption{Forecasted uncertainties on parameters of interest, and bias on $f_{\rm NL}$ due to neglecting relativistic effects from number counts and kSZ measurements. Forecasts are shown with and without alignment bias, and as additional observables and priors are added. }
\label{table:constraints}
\end{table}

We begin by examining constraints on the various bias functions in Fig.~\ref{fig:bias_constraints}.
These can be seen to drastically deteriorate as alignment bias is taken into
account (compare the gray-dashed curves with the others), especially galaxy
bias with which the alignment bias is degenerate. The bias functions
subsequently recover as information from the CMB is included in the constraint.
Notably, beyond the primary CMB and number counts alone, kSZ tomography can offer
a substantial improvement in constraints on these biases, from a factor of a few, up to an order of magnitude in
the case of evolution bias (for $z \lesssim 2.5$). No priors on any parameters are included in the constraints in this figure,
and cosmological parameters including $f_{\rm NL}$ and the $\epsilon_{\rm GR}$ parameters are marginalized over.

The fiducial values of the galaxy and magnification bias as described in Appendix~\ref{appendix:numbercounts}
are of order unity, and thus should be detected at high significance. The evolution bias is somewhat smaller, so detection
prospects are perhaps marginal even when including information from kSZ tomography in the constraint.
The evolution bias itself is degenerate with the amplitude of relativistic effects, and to a smaller extent $f_{\rm NL}$.
An improved constraint on this bias can therefore help to reduce forecasted uncertainties,
especially in $\epsilon_{\rm GR}^{\rm N}$. The amplitude of the alignment bias, on the other hand,
is expected to be at most a few percent \cite{0903.4929}; while also marginal, we find that kSZ
tomography may offer a way to detect alignment bias. The bias due to optical
depth degeneracy, $b_v$,
is only constrained when including information from the remote dipole field. The sub-percent
constraint on $b_v$ is highly significant relative to the fiducial value of unity of this
bias, and is comparable to constraints forecasted using other methods \cite{1901.02418}.

We next examine how detectable relativistic contributions to observables
are, looking at $\epsilon_{\rm GR}^{\rm N}$ and $\epsilon_{\rm GR}^{\rm kSZ}$,
along with how an inferred bias for $f_{\rm NL}$ may be incurred when neglecting
relativistic effects. The constraints and bias values are listed in Table~\ref{table:constraints}.
These have been marginalized over other cosmological parameters noted in
Table~\eqref{table:params_list}, as well as all bias functions, including the bias due
to the kSZ optical depth degeneracy where relevant. In both the case where we include and neglect alignment bias,
relativistic effects are not detectable with number counts alone, and the bias due to neglecting these effects is
unimportant. As constraints from the primary CMB are added, which primarily serve to pin
down standard cosmological parameters, constraints on relativistic effects
recover from alignment bias. As the reconstructed kSZ
remote dipole field is included in the analysis, the constraints improve substantially,
to the point where both a bias on $f_{\rm NL}$ and relativistic effects are detectable
at moderate significance.

Following \cite{1505.07596}, we also consider the impact of including a
Gaussian prior on the evolution bias
of $\Delta f_{\rm evo} = 1$ in all redshift bins. The constraints on relativistic effects
we find are then comparable to the multiple tracers considered in \cite{Alonso:2015sfa},
however the prior we use is significantly less strict.
We additionally consider what happens when we do not simultaneously constrain the $\epsilon_{\rm GR}$
parameters and $f_{\rm NL}$: this improves constraints to $\sigma(f_{\rm NL})
\sim 0.9$ both with and
without the evolution bias prior.

\begin{figure}[htb]
  \centering
      \includegraphics[width=0.99\textwidth]{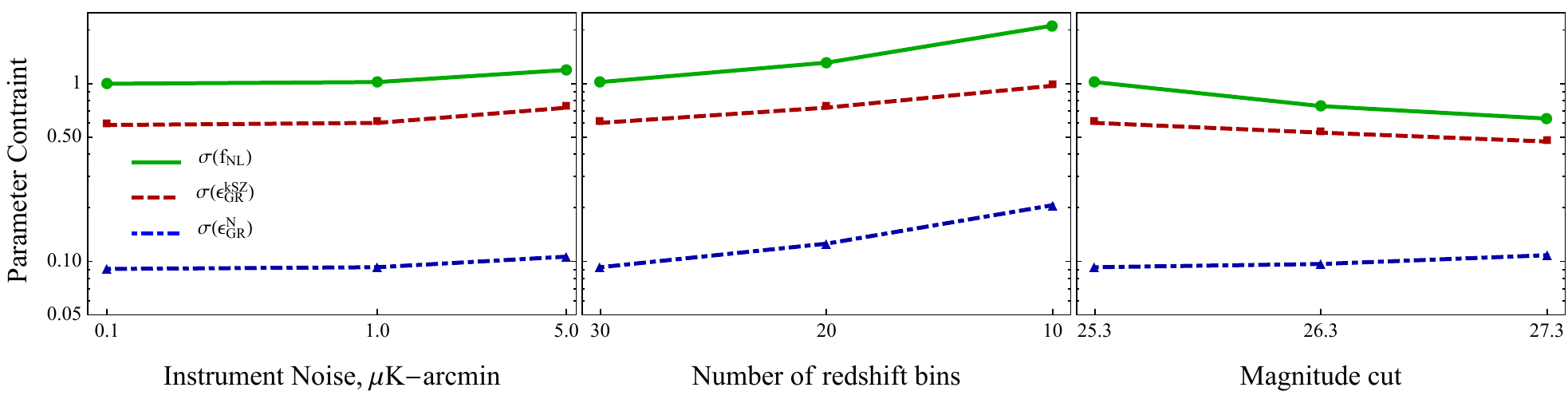}
    \caption{
      Forecasted uncertainty in parameters as experimental parameters are varied.
      Fiducial parameters include 1 $\mu$K-arcmin noise, 30 bins, and a
      magnitude limit of $r=25.3$ (magnitude limit discussed in more detail in
      Appendix~\ref{appendix:numbercounts}).
    }
  \label{fig:parameter_constraints}
\end{figure}

Using all tracers considered--number counts including alignment bias and the $f_{\rm evo}$ prior,
the primary CMB, and the remote dipole field--we lastly explore the
constraints as a function of instrumental parameters. We show these in Figure~\ref{fig:parameter_constraints},
as the CMB experiment noise, number of redshift bins (ie. photometric redshift error), and limiting magnitude are
varied about the fiducial values considered above. The constraints presented in these figures include the evolution bias prior.

Generally, the constraints improve as one might expect. Increasing CMB instrument noise results in
higher reconstruction noise on the remote dipole field, and constraints on the parameters
we consider all become mildly worse. Decreasing the number of redshift bins
(e.g. increasing the photometric redshift error) similarly reduces the information available,
resulting again in moderately worse constraints. Varying the limiting
magnitude of the galaxy survey provides a more complicated picture: constraints on $f_{\rm NL}$
and $\epsilon_{\rm GR}^{\rm kSZ}$ improve, while constraints on $\epsilon_{\rm GR}^{\rm N}$ actually
worsen. This is because the fiducial bias model changes in such a way that the signal of relativistic effects
(as well as lensing, which we do not consider) all decrease relative to intrinsic density perturbations
and RSD effects. The ``noise'' due to cosmic variance from these dominant effects is therefore
increased, resulting in a worse overall constraint on relativistic effects.

\section{Conclusions and discussion}
\label{sec:discussion}

In this paper, we have illustrated the potential importance of the remote dipole field reconstructed using kSZ tomography
for the detectability of relativistic effects in galaxy number counts, and the need to account for relativistic effects
to obtain an unbiased measurement of primordial non-Gaussianity using kSZ tomography. We also highlighted
the improvement on the measurement of various (redshift-dependent) bias parameters from the correlation between the remote dipole field
and number counts. The forecasted constraints on most bias parameters can improve by a factor of a few, and constraints
on the evolution bias can improve by more than an order of magnitude. This is significant, as evolution bias is
strongly degenerate with both relativistic effects and scale-dependent bias from primordial non-Gaussianity. Even with
this improvement, evolution bias is poorly constrained; we demonstrate that a weak prior can substantially improve the detectability
of relativistic effects. Importantly, our analysis shows that kSZ tomography can significantly mitigate the effects of alignment bias,
which can seriously degrade constraints on primordial non-Gaussianity using number counts alone.

Our forecasts have included a number of optimistic assumptions. In particular, we have presented the constraints
in the limit where we have data on the full sky and foregrounds and systematics are negligible. Partial-sky data will
weaken measurements on the largest scales, which could significantly degrade constraints on relativistic effects and
primordial non-Gaussianity. Foregrounds and systematics on small angular scales have the potential to degrade the
reconstruction of the remote dipole field. To estimate the effect this might have, we repeated our forecast using an $\ell_{\rm max}$
of 5000 in the reconstruction noise. This yields only a roughly 25\,\% increase in our forecasted uncertainties.

Several additional assumptions made in this work are perhaps pessimistic. For example,
the galaxy number counts we compute contain very few observable galaxies beyond redshift $z \sim 2$,
while calculations of the galaxy number counts using different assumptions can yield
 a higher number of galaxies at comparable redshifts (see e.g.~\cite{Smith:2018bpn,Schmittfull:2017ffw} for similar studies).
The magnitude limits we assume are also conservative, and information from fainter galaxies up to a magnitude of
$r \sim 27.5$ may be accessible, albeit with larger uncertainties in photometric redshifts.
We have also not split our analysis into separate galaxy
populations (red, blue) nor included additional tracers, such as intensity mapping, that have been shown
to improve constraints~\cite{1505.07596,1710.02477,Alonso:2015sfa}.

While changing our assumptions about the fiducial CMB and galaxy surveys has the potential to nudge our
forecasted uncertainties up or down, this work demonstrates that kSZ tomography promises to be
an important and useful tool for isolating both relativistic effects and new physics.
We have further highlighted a significant bias on measurements of
primordial non-Gaussianity incurred when relativistic effects are neglected, and illustrated a significant improvement
on various bias functions that can be obtained using kSZ tomography. This paper strengthens the science case
for performing kSZ tomography using future observations.

\section{Acknowledgments}
We would like to thank Juan Cayuso, Neal Dalal, and Moritz Munchmeyer for helpful discussions.
This research was enabled in part by support provided by the Shared Hierarchical Academic Research
Computing Network (SHARCNET:www.sharcnet.ca), Compute Canada (www.computecanada.ca), and the Kenyon
College Department of Physics. This research was supported in part by Perimeter Institute for
Theoretical Physics. Research at Perimeter Institute is supported by the Government of Canada through
the Department of Innovation, Science and Economic Development Canada and by the Province of Ontario
through the Ministry of Research, Innovation and Science. MCJ and DC were supported by the National
Science and Engineering Research Council through a Discovery grant. JBM acknowledges support as a
CITA National Fellow.

\bibliographystyle{utcaps}
\bibliography{references}

\appendix

\section{Number counts transfer functions}
\label{appendix:numbercounts}

Corrections to relativistic perturbations arise when considering how to map from the
observed $\left(N(z)-\bar{N}(z)\right)/\bar{N}(z)$ (or the unperturbed $N(z)$)
to the theoretical $\left(N(\bar{z})-\bar{N}(\bar{z})\right)/\bar{N}(\bar{z})$.
Here we describe the standard linear theory equations used to compute the
perturbations in number counts, and the adjustments we make to these equations in order
to account for non-Gaussianities and alignment bias.

We define all number counts transfer functions used below. These transfer functions are the
same as can be found in literature \cite{1105.5292,1105.5280,Alonso:2015sfa,1505.07596},
with two amendments: two additional bias functions are
included, one for alignment bias, $b_{\rm A}$ \cite{0903.4929},
and one for primordial non-Gaussianities, $b_{\rm NG}$ \cite{Dalal:2007cu}.
\begin{align}
\label{eq:ngr_transfers}
\Delta_{\ell}^{{\rm D},i}(k) & =\int d\chi \tilde{W}_{i}(\chi)\,(b_{\rm G}-b_{\rm A}/3 + b_{\rm NG})\,\,S_{\delta_M,{\rm syn}}(k,\chi)\,j_{\ell}(k\chi) \\
\Delta_{\ell}^{{\rm RSD},i}(k) & =\int d\chi \tilde{W}_{i}(\chi)\,(1+b_{\rm A}/f)\frac{k^2 S_v(k,\chi)}{aH}j_{\ell}''(k\chi) \\
\Delta_{\ell}^{{\rm lens},i}(k) & =\ell(\ell+1)\int d\chi \tilde{W}_{i}(\chi) \int_0^{\chi} d\chi' \frac{2-5s(\chi')}{2} \frac{\chi-\chi'}{\chi\chi'}S_{\phi+\psi}(k,\chi')j_{\ell}(k\chi') \\
\Delta_{\ell}^{{\rm D1},i}(k) & = \int d\chi \tilde{W}_{i}(\chi) \left(\frac{\mathcal{H}'}{\mathcal{H}^2} + \frac{2-5s}{\chi\mathcal{H}} + 5s - f_{\rm evo} \right) k S_v(k,\chi) j_\ell'(k\chi) \\
\Delta_{\ell}^{{\rm D2},i}(k) & = \int d\chi \tilde{W}_{i}(\chi) (f_{\rm evo} - 3) \mathcal{H} S_v(k,\chi) j_\ell(k\chi) \\
\Delta_{\ell}^{{\rm P1},i}(k) & = \int d\chi \tilde{W}_{i}(\chi) \left(\frac{\mathcal{H}'}{\mathcal{H}^2} + \frac{2-5s}{\chi\mathcal{H}} + 5s - f_{\rm evo} + 1 \right) S_\psi(k,\chi) j_\ell(k\chi) \\
\Delta_{\ell}^{{\rm P2},i}(k) & = \int d\chi \tilde{W}_{i}(\chi) (-2+5s) S_\phi(k,\chi) j_\ell(k\chi) \\
\Delta_{\ell}^{{\rm P3},i}(k) & = \int d\chi \tilde{W}_{i}(\chi) \frac{1}{\mathcal{H}} S_{\phi'}(k,\chi) j_\ell(k\chi) \\
\Delta_{\ell}^{{\rm P4},i}(k) & = \int d\chi \tilde{W}_{i}(\chi) \left( \frac{\mathcal{H}'}{\mathcal{H}^2}+\frac{2-5s}{\chi\mathcal{H}} +5s - f_{\rm evo} \right) \int_0^{\chi} d\chi' S_{(\phi+\psi)'}(k,\chi') j_\ell(k\chi') \\
\Delta_{\ell}^{{\rm P5},i}(k) & = \int d\chi \tilde{W}_{i}(\chi) \frac{1}{\chi} \int_0^{\chi} d\chi' (2-5s(\chi')) S_{\phi+\psi}(k,\chi') j_\ell(k\chi')\,,
\label{eq:p5term}
\end{align}
where $\chi$ is comoving distance and the redshift-space window function
\begin{equation}
\label{eq:ngr_windowfn}
W_{i}(z) \equiv \left| \frac{d\chi}{dz} \right| \tilde{W}_i(\chi)
\end{equation} is a tophat function in
redshift, i.e., nonzero and constant in the i-th redshift bin, and
is normalized so that its integral over redshift is unity. The transfer functions
for determining various fields from primordial perturbations are noted by $S_{\delta_M,{\rm syn}}$
for the synchronous gauge matter density, $S_{\phi}$ and $S_{\psi}$ for the Newtonian
potentials, and $S_{v}$ for the scalar velocity field as defined in \cite{Terrana2016},
which is related to the divergence of the velocity field $\theta$ as $S_{\theta} = k^2 S_{v}$.
The bias due to non-Gaussianities (NG) is described by
\begin{equation}
b_{\rm NG}(k, z) = 3 f_{\rm NL} (b_{\rm G}(k, z) - 1) \Omega_m H_0^2 \delta_c / ( k^2 T(k) S_{\psi} )\,,
\end{equation}
with $\delta_c = 1.686$ the linearized collapse threshold.

The remaining bias functions introduced in
Eqs.~\eqref{eq:ngr_transfers}--\eqref{eq:p5term} are defined with respect to
the \emph{background} source population. Specifically we can take
Eq.~\eqref{eq:ngr_transfers} as the definition of the galaxy bias ($b_G$), and
use the parameterization (for a full galaxy sample) based on the simulations of
\cite{Weinberg2002} and  quoted in the LSST science book \cite{0912.0201}, of
$b_G = 0.95 / D(z) \simeq 0.95 + 0.67z$. The magnification bias ($s$) and
evolution bias ($f_{\rm evo}$) are defined as
\begin{align}
  s(\chi) &\equiv \frac{5}{2} \frac{n_s(\chi, \ln{L_{\rm
  cut}})}{\mathcal{N}(\chi, > \ln{L}_{\rm cut})}, \\
  f_{\rm evo}(\chi) &\equiv \frac{\partial \ln{a^3\mathcal{N}(\chi, >
  \ln{L}_{\rm cut})}}{\partial \ln{a}}.
  \label{eq:bias_functions}
\end{align}
Here, the quantity $n_s$ is the luminosity function: the galaxy number counts density
per luminosity on a spatial hypersurface, ie. not projected onto a lightcone,
\begin{equation}
  n_s \equiv \frac{\rm \#\,galaxies}{dV d\ln L},
\end{equation}
and $\mathcal{N}$ is the integrated number density above
a threshold luminosity, derived from the luminosity function,
\begin{equation}
  \mathcal{N}(\chi, > \ln L_{\rm cut}) \equiv \int_{\ln L_{\rm cut}}^\infty d
  \ln L' n_s(\chi, \ln L').
  \label{eq:nfunctions}
\end{equation}
We model the evolution and magnification biases following the approach
outlined in Appendix B.4 of \cite{1505.07596}. Explicitly, we assume a
Schechter luminosity function of the form
\begin{align}
  n_s(M) dM &= 0.4 \ln(10) \phi_* \left[10^{0.4(M_*-M)}\right]^{\alpha + 1}
  \exp{\left[-10^{0.4(M_*-M)}\right]} dM\,.
  \label{eq:schechter}
\end{align}
To model the full sample of galaxies we use the $r'$-band luminosity
function found by \cite{Gabasch2005} to approximate the $r$ band luminosity
function,
\begin{align}
  M_*(z) &= M_0 + g \ln(1+z), \\
  \phi_*(z) &= (\phi_0 + \phi_1 z + \phi_2 z^2) ~ [10^{-3}\,\mathrm{Mpc}^{-3}].
  \label{eq:gabaschlum}
\end{align}
The parameters of this model have the explicit values: $\alpha = -1.33$, $g = -1.25$, $M_0 = -21.49$,
$\phi_0 = 2.59$, $\phi_1 = -0.136$, and $\phi_2 = -0.081$.

We then relate the absolute magnitude in Eq.~\eqref{eq:schechter} to an
apparent magnitude, $m$, via
\begin{align}
  M &= m - 25 - 5 \log_{10}\left[\frac{d_L(z)}{1\, \mathrm{Mpc}\, h^{-1}}\right] +
  \log_{10} h - k(z),
  \label{eq:magdist}
\end{align}
where the $d_L$ is the luminosity distance and $k$ is a $k$-correction due to
the corresponding galaxy's spectral energy distribution redshifted to $z$. We
use the extrapolated values of $k(z)$ found in \cite{1505.07596} (with $k(z)
\propto z$). The quantities, $s$ and $f_{\rm evo}$ (and their redshift
dependence) will thus depend on the magnitude limit of the survey in question
(we need to integrate Eq.~\eqref{eq:nfunctions} to $M_{\rm cut}$). Here we use
the limit $m_{\rm cut} = r = 25.3$ as a representative choice for an
LSST type survey \cite{0912.0201}, although we also explore $m_{\rm cut} = r = \,26.3,\,27.3$.

Lastly, and explicitly, we consider the general relativistic terms to include
\begin{equation}
\Delta^{{\rm N,GR},i}_{\ell} = \Delta^{{\rm D1},i}_{\ell} + \Delta^{{\rm D2},i}_{\ell} + \Delta^{{\rm P1},i}_{\ell} + \Delta^{{\rm P2},i}_{\ell} + \Delta^{{\rm P3},i}_{\ell} + \Delta^{{\rm P4},i}_{\ell} + \Delta^{{\rm P5},i}_{\ell}\,.
\end{equation}
The remaining lensing, RSD, and intrinsic perturbation terms we do not consider as part
of the relativistic effects. Together, both the relativistic and non-relativistic
contributions comprise the first-order gauge-independent observable angular power
spectrum.

\section{Remote dipole (kSZ) transfer functions}
\label{appendix:ksz}

The contributions to the remote dipole field transfer function are given as
\begin{equation}
\Delta_{\ell}^{A,i}(k) = \frac{b_{v}}{2\ell+1} \int d\chi\,W_{i}(\chi) [ \ell j_{\ell-1}(k \chi)
  - (\ell + 1) j_{\ell+1}(k \chi) ] \mathcal{K}^A(\chi)
\end{equation}
where $b_v$ is an overall bias in the amplitude of the reconstructed remote dipole field
that arises due to uncertainty in the electron density field, the ``optical depth bias'',
and the kernels $\mathcal{K}^A$ are given by one of
\begin{align}
\mathcal{K}^{\rm LD}(\chi) & =  - k S_v(k,\chi) \\
\mathcal{K}^{\rm RD}(\chi) & =  k S_v(k,\chi_{\rm CMB}) [ j_0(k\Delta\chi) - 2j_2(k\Delta\chi) ] \\
\mathcal{K}^{\rm SW}(\chi) & =  3 \left( 2 S_\psi(k,\chi) - \frac{3}{2} \right) j_1(k\Delta\chi) \\
\mathcal{K}^{\rm ISW}(\chi) & =  6 \int_{\chi_{\rm CMB}}^{\chi} d\chi' \frac{dS_{\psi}(k,\chi')}{d\chi'} j_1(k\Delta\chi')
\end{align}
for the various terms $A$ (local Doppler LD, remote Doppler RD, Sachs-Wolfe SW, and integrated Sachs-Wolfe ISW) that contribute to the remote dipole field, and for $\Delta\chi \equiv \chi - \chi_{\rm CMB}$.
The fiducial optical depth bias value is chosen to be unity, $b_v = 1$, and is marginalized over
independently in each redshift bin considered.
Explicitly, we consider the general relativistic (or primary CMB) terms to include
\begin{equation}
\Delta^{{\rm kSZ,GR},i}_{\ell} = \Delta^{{\rm RD},i}_{\ell} + \Delta^{{\rm SW},i}_{\ell} + \Delta^{{\rm ISW},i}_{\ell}\,.
\end{equation}
The non-relativistic (non-primordial) remaining term is the local Doppler contribution,
attributable to only the Newtonian-gauge peculiar velocity of the remote electron.
As noted in \cite{Terrana2016}, the Newtonian velocity contribution alone is not a
(linear) gauge-invariant quantity, and the observable should include the
additional relativistic contributions.

\end{document}